\def\revtex{1}

\ifx\revtex\undefined

\documentclass[entropy,article,submit,oneauthor,pdftex]{Definitions/mdpi}

\firstpage{1}
\pubvolume{1}
\issuenum{1}
\articlenumber{0}
\pubyear{2021}
\copyrightyear{2021}
\datereceived{}
\dateaccepted{}
\datepublished{}
\hreflink{https://doi.org/} 

\Title{On The Complete Description Of Entangled Systems\\
Part I: Hidden Variables And The Context Communication Cost of Simulating Quantum Correlations}

\TitleCitation{The Context Communication Cost of Simulating Quantum Correlations}

\Author{Karl Svozil $^{1}$\orcidA{}}

\AuthorNames{Karl Svozil}

\AuthorCitation{Svozil, K.}

\address[1]{%
$^{1}$ \quad Institute for Theoretical Physics, TU Wien, Wiedner Hauptstrasse 8-10/136, 1040 Vienna,  Austria; svozil@tuwien.ac.at; \url{http://tph.tuwien.ac.at/~svozil}}

\corres{Correspondence: svozil@tuwien.ac.at}

\abstract{Some forms of classical simulations of quantum type probabilities and correlations are capable of violating Boole's conditions of possible experience, such as the Clauser-Horne-Shimony-Holt inequality, even beyond the Tsirelson bound. This can be achieved by communicating a single bit encoding the measurement context.}


\DeclareFontFamily{U}{bbold}{}
\DeclareFontShape{U}{bbold}{m}{n}
 {
  <-5.5> s*[1.069] bbold5
  <5.5-6.5> s*[1.069] bbold6
  <6.5-7.5> s*[1.069] bbold7
  <7.5-8.5> s*[1.069] bbold8
  <8.5-9.5> s*[1.069] bbold9
  <9.5-11> s*[1.069] bbold10
  <11-15> s*[1.069] bbold12
  <15-> s*[1.069] bbold17
 }{}

\begin{document}

\else
\documentclass[%
      reprint,
   twocolumn,
 amsmath,amssymb,
 aps,
 pra,
  longbibliography,
  floatfix,
 ]{revtex4-2}

\usepackage[dvipsnames]{xcolor}

\usepackage{amssymb,amsthm,amsmath,bm}

\usepackage{tikz}
\usetikzlibrary{angles, arrows.meta, positioning, quotes, decorations.pathmorphing, decorations.pathreplacing, decorations.shapes,calc,decorations.pathreplacing,decorations.markings,positioning,shapes,snakes}

\usepackage[breaklinks=true,colorlinks=true,anchorcolor=blue,citecolor=blue,filecolor=blue,menucolor=blue,pagecolor=blue,urlcolor=blue,linkcolor=blue]{hyperref}
\usepackage{url}

\ifxetex
%
%
\usepackage{fontspec}
\usepackage{fontspec}
\setmainfont{Garamond}
\setsansfont{Garamond}
\fi

\usepackage{mathbbol} 

\begin{document}

\title{On The Complete Description Of Entangled Systems\\
Part I: Hidden Variables And The Context Communication Cost of Simulating Quantum Correlations}

\author{Karl Svozil}
\email{svozil@tuwien.ac.at}
\homepage{http://tph.tuwien.ac.at/~svozil}

\affiliation{Institute for Theoretical Physics,
TU Wien,
Wiedner Hauptstrasse 8-10/136,
1040 Vienna,  Austria}

\date{\today}

\begin{abstract}
Some forms of classical simulations of quantum type probabilities and correlations are capable of violating Boole's conditions of possible experience, such as the Clauser-Horne-Shimony-Holt inequality, even beyond the Tsirelson bound. This can be achieved by communicating a single bit encoding the measurement context.
\end{abstract}


\maketitle

\fi

\section{The Einstein-Podolsky-Rosen (EPR) Conundrum}

To properly appreciate the difference between quantum entanglement and classical correlations, we shall construct classical states
that perform certain features of quantum entangled states:
they have value indefinite (individual) outcomes yet are correlated or relationally encoded such that,
measurement of a particular outcome on one side entails certainty of the outcome of the same experiment on the other side.
The quantum-classical difference lies in the ontology of the microstates:
wheres the classical physical state is a statistical one with totally specified individual value definite properties---in
principle giving rise to an infinity of value definite individual outcomes---such
a value definiteness in quantum mechanics is postulated to be true for only one particular context
(or a ``star-shaped'' multiplicity of contexts~\cite{2012-incomput-proofsCJ}) corresponding to the particular pure state the particle was prepared in (preselected).

\section{EPR type configurations with classical shares}

In the following section, quasi-classical models of singlet states are studied.
It turns out that correlations may be perfect, just like in the quantum case.
And just like in the quantum case, the classical constraints on correlations as a function of the differences between microstates can but may not be linear.

However, the stochastic behavior of their micro- versus macro-states is very different:
Whereas insistence on individual micro-states results in deterministic outcomes, bundling of such microstates effectively yields value indefiniteness
of the macro-states.

This micro- versus macro-state difference grounded in epistemology versus epistemology has its correspondence in classical statistical mechanics~\cite{Myrvold2011237}.
In Maxwell's own words~\cite[p.~442]{garber},
{\em ``I carefully abstain from asking the molecules which enter where they last started
from. I only count them and register their mean velocities, avoiding all personal
inquiries which would only get me into trouble.''}
Hence~\cite[p. 279]{Maxwell-1878},
{\em ``The truth of the second law is, therefore, a statistical,
 not a mathematical, truth, for it depends on the fact that the bodies we deal with consist
of millions of molecules, and that we never can get hold of single molecules.''}

\subsection{Peres' bomb fragment model as classical mechanics ``singlet'' analogon}

For the sake of a concrete classical simulacrum, we first review an explanatory article~\cite{peres222} in which
Peres introduced the model of a
{\em ``bomb, initially at rest} [[with zero angular momentum, that]] {\em explodes into two fragments carrying opposite angular momenta''.}
Dichotomic observables of those individual fragments are then defined by
$\mathbf{r}_{\boldsymbol{\alpha}} = \mathrm{sign} \left( \boldsymbol{\alpha} \cdot \mathbf{J} \right)$,
where $\boldsymbol{\alpha}$ is a unit vector in an arbitrary direction, chosen by the observer,
and $\mathbf{J}$ is that individual particle's angular momentum.

Disregarding other physical categories and features we may call $\mathbf{J}$ the ``state'' of the explosion fragments.
Thereby, ontological realism implicitly assumes that $\mathbf{J}$ exists independent of any observing mind~\cite{stace}.
A precise specification of $\mathbf{J}$ may require an infinite amount of information; e.g., in terms of some orthonormal basis,
the specification of that basis, as well as the respective three real-valued coordinates.

The epistemology of this classical configuration needs some clarification:
on the one hand, $\mathbf{J}$ can---at least in principle---be measured and operationalized to arbitrary precision;
for instance, by probing it with the dichotomic observables $\mathbf{r}_{\boldsymbol{\alpha}}$
at some arbitrary angle $\boldsymbol{\alpha}$.
The means invested and the precision obtained is thereby due to a particular choice of the experimenter.
Classically one could even pretend to be able to control the outcome of the explosion---and thus produce an exactly specified state
not at random but at will~\cite{diaconis:211}.

On the other hand, for the sake of the analogy to quantum EPR configurations,
and for all practical purposes and means the parameter $\mathbf{J}$
shall be assumed to remain effectively hidden and unknown to the experimenter
before measuring the observable $\mathbf{r}_{\boldsymbol{\alpha}}$.
Moreover, the experimenter is assumed to be effectively (that is, with the means employed)
unable to control or choose the individual particle's angular momentum $\mathbf{J}$.
This is a reasonable epistemic consequence of the supposedly uncontrollable detonation of the bomb nested in its environment,
even if ontological realism claims otherwise.
Therefore,  for this practical purpose, $\mathbf{J}$ shall be assumed to be equidistributed over all spatial directions;
in Peres' terms, ``unpredictable and randomly distributed''.

Note that, by construction, two experimenters share individual fragments of the bomb.
Because of angular momentum conservation (and previous angular momentum zero), each fragment carries the same amount of angular momentum information
in ``opposing'' states $\mathbf{J}$ and $-\mathbf{J}$.

As a consequence of this situation,
and from the intrinsic, endogenous, private, subjective view of each one of the two experimenters $\mathbf{A}$ and $\mathbf{B}$,
the observed respective outcome $\mathbf{r}_{\boldsymbol{\alpha}}$ or $\mathbf{r}_{\boldsymbol{\beta}}$ of any single observation
appears uncontrollable and unpredictable.
Because relative to the means available to $\mathbf{A}$ and $\mathbf{B}$, the parameter $\mathbf{J}$ is effectively hidden.

It thus (wrongly) seems as if the outcomes on each one of the two sides occur irreducibly randomly.
They appear to be ``mysteriously'', spontaneously and continuously created
(in older theological terminology, {\em Creatio Continua} or {\em Ongoing Creation})
contextually through nesting with the environment of the measurement apparatus on (possibly spatially separated)
sides  $\mathbf{A}$ and $\mathbf{B}$;
akin to
Bohr's~\cite{bohr-1949,Khrennikov2017,Jaeger2019} conditionality of phenomena, a contingency due to
{\em ``the impossibility of any sharp separation between the behavior of atomic
objects and the interaction with the measuring instruments which serve to define the conditions
under which the phenomena appear.''}

And yet, joint outcomes from the same explosion are (cor)related:
statistically, that is, for many experimental runs and averaging over many outcomes,
a straightforward geometric argument~\cite{peres222} yields a correlation
$
\langle \mathbf{r}_{\boldsymbol{\alpha}} \mathbf{r}_{\boldsymbol{\beta}}\rangle
= 2 \theta / \pi - 1
$
linear in the angle $\theta = \angle{ \boldsymbol{\alpha} \boldsymbol{\beta} }$ between $\boldsymbol{\alpha}$ and $\boldsymbol{\beta}$.

For individual pairs of outcomes associated with two fragments of the same bomb,
it is always the case that, if both experimenters $\mathbf{A}$ and $\mathbf{B}$ measure the same observable $\mathbf{r}_{\boldsymbol{\alpha}}$---that is,
$\boldsymbol{\beta}=\boldsymbol{\alpha}$, then these observers end up with exactly inverse events:
if $\mathbf{A}$ obtains outcome $\mathbf{r}_{\boldsymbol{\alpha}} = \pm 1$ then $\mathbf{B}$ obtains outcome $-\mathbf{r}_{\boldsymbol{\alpha}} = \mp 1$
along the same direction $\boldsymbol{\alpha}$ (and vice versa).
This is due to the configuration (original angular momentum zero) and conservation of momentum:
because if $\mathbf{A}$ measures $\pm \mathbf{J}$ then $\mathbf{B}$ measures $\mp \mathbf{J}$ (and vice versa).

This relational property just described is independent of the spatio-temporal distance of the events.
In particular, the events can be spatially separated under strict Einstein locality conditions,
such that no causal communication can occur between the two observers $\mathbf{A}$ and $\mathbf{B}$.
These observers might find it mindboggling that,
although ``far away'' and causally (relativistically) separated,
and although any single event occurs seemingly irreducibly random on their respective sides,
the state is relationally encoded such that, every random outcome at one side entails the exact opposite random outcome on the other side.

This apparent mindboggling feature resolves that the apparent randomness is an intrinsic illusion,
because extrinsically---that is, from an ontologic perspective---both (spatially) separated observers work on the same shared state $\mathbf{J}$.
This unknown share is both the ``source of (intrinsic) randomness'' and of the relational encoding.
It is important to realize that in this case the share and the two fragments representing it is value definite and precisely defined.
Randomness here is epistemic, and, contrary to Bohr's suggestion of contextuality,
neither the environment nor any type of Wigner's friend nesting contributes to the outcome.
One may also say that the interface or Heisenberg cut is located at the share, that is, at the individual fragments.
Those fragments, to emphasize and repeat, carry the share $\mathbf{J}$ and are value definite.

\subsection{Finite set partitions as singlet states}

One feature of Peres' model discussed earlier is the infinite amount of information necessary to specify the ``hidden parameter'' share $\mathbf{J}$.
In what follows a finite quasi-classical set representable partition model will be presented that allows similar EPR-type considerations but delineates the
basic assumptions even further. It is based on partition logics~\cite{schaller-96,schaller-95,svozil-2001-eua} that are pasting of blocks~\cite{nav:91}
many allowing a faithful orthogonal representation~\cite{svozil-2018-b} by a vertex labeling of vectors
that are mutually orthogonal within blocks~\cite{lovasz-79,Portillo-2015}.
For the sake of comparison to the Clauser-Horne-Shimony-Holt (CHSH) configuration, consider again pairs of particles with four potential observables
$\mathbf{a} , \mathbf{a}' , \mathbf{b} , \mathbf{b}'$.
Suppose that $\mathbf{a} , \mathbf{a}$ are measured on experimenter $\mathbf{A}$'s side, and
$\mathbf{b} , \mathbf{b}$ are measured on experimenter $\mathbf{B}$'s side, respectively.

For the convenience of comparison with the CHSH configuration we again suppose that these observables are dichotomic, with outcomes $+1$ or $-1$;
that is, more explicitly, $\mathbf{a} , \mathbf{a}' , \mathbf{b} , \mathbf{b}' \in \{-1,+1\}$.
Suppose further that we are forming ``singlets'' by ``bundling opposite-valued'' particle pairs, represented as ordered pairs
$\mathbf{H}=\big( \{\mathbf{a} , \mathbf{a}' , \mathbf{b} , \mathbf{b}' \} , \{-\mathbf{a} , -\mathbf{a}' , -\mathbf{b} , -\mathbf{b}' \} \big)$
of two four-tuples per singlet state, forming the singlet states.
There are $2^4=16$ different types of pairs or singlet states, namely, in lexicoraphic order ($-1 < +1$),
\begin{center}
\begin{tabular}{l}
$\big[ \{-1 , -1 , -1 , -1 \} , \{1,1,1,1, \} \big]$,\\
$\big[ \{-1 , -1 , -1 , 1 \} , \{1,1,1,-1, \} \big]$,\\
$\qquad \ldots$                                           \\
$\big[ \{1,1,1,-1\} , \{-1 , -1 , -1 , 1 \} \big]$, and\\
$\big[ \{1,1,1,1, \} , \{-1 , -1 , -1 , -1 \} \big]$.
\end{tabular}
\end{center}
A generalization to more observables is straightforward.

Suppose further that we are filling a generalized urn~\cite{wright} with such ordered pairs, and draw (choose) them ``at random''.
We might imagine these ordered pairs as two balls painted uniformly black; printed on this black background
are the symbols ``$-$'' (for value $-1$) or ``$+$'' (for value $+1$) in exactly
four mutually different colors---one color for each one of the four observables $\mathbf{a} , \mathbf{a}' , \mathbf{b} , \mathbf{b}'$.
Suppose that
experimenter $\mathbf{A}$ wears two types of eye-glasses, making visible either observable $\mathbf{a}$ or observable $\mathbf{a}'$;
likewise, experimenter $\mathbf{B}$ wears two other types of eye-glasses, making visible either observable $\mathbf{b}$ or observable $\mathbf{b}'$.
Suppose that,  in this subtractive color scheme, all other three colors appear black.
Each one of the two observers sees exactly one of the four observables.

The ``hidden parameters'' in this case are the values of $\mathbf{a} , \mathbf{a}' , \mathbf{b} , \mathbf{b}'$
in the full or complete state or share $\mathbf{H}$; and yet the way this model or game is constructed,
every single experimental run accesses only two of them, corresponding to the choices $\mathbf{a}$ versus $\mathbf{a}'$,
and $\mathbf{b}$ versus $\mathbf{b}'$.
(Of course, experimenters may cheat and put off their eyeglasses, thereby seeing the full state with all variations.)

Classical probabilities and expectations of such models can be obtained by forming the convex sum over all extreme cases or states.
In particular, the CHSH bounds to these probabilities and expectations are obtained by
(i) first interpreting the tuples codifying these 16 extreme cases or two-valued states as vector coordinates
(with respect to an orthonormal basis such as the Cartesian standard basis),
(ii) then consider the convex polytope defined by identifying these 16 vectors as vertices of the polytope (in four-dimensional vector space $\mathbb{R}^4$), and
(iii) finally solving the hull problem, yielding the hull inequalities
characterizing the ``inside-outside'' borders of this convex polytope~\cite{froissart-81,pitowsky-86,Pit-94,svozil-2017-b}.

As mentioned earlier, we must make a distinction between ontology versus epistemology, or, in another conceptualization,
extrinsic versus intrinsic, operational observables.
In this case, it is very transparent that the seemingly random occurrence of outcomes originated in the choice or draw of the pair from the (generalized) urn.
There is no influence of the environment on the outcome, and thus contextuality in the way imagined by Bohr is absent.

\subsection{Microstates versus macrostates}

The value indefiniteness of experimental outcomes in both classical cases resides in the ignorant choice of the particular share:
in the Peres model $\mathbf{J}$, the angular momentum vectors of the fragments, and for the discrete generalized urn model in the random
selection of the particular elements of the set $\mathbf{H}$ of ball pairs.
Insofar as we are unwilling or incapable of observing those individual micro-states directly we can speak of the
``irreducible randomness'' of individual outcomes; but this irreducibility is based on epistemic limitations
to ``grab'' the individual share and consider it as the microstate.

Once the microstate is specified, the correlation for identical measurements is perfect.
For non-identical measurements, the share  $\mathbf{J}$ yields linear correlations
$
\langle \mathbf{r}_{\boldsymbol{\alpha}} \mathbf{r}_{\boldsymbol{\beta}}\rangle
= 2 \theta / \pi - 1
$
as a function of the (relative) angle $\theta$.
(in the plane normal to the fragment's velocity).
The correlations in the discrete urn case are bound by linear constraints
called ``conditions of possibly experience'' by Boole~\cite{Boole,Pit-94};
four of these constraints are called CHSH inequality for historical reasons.


\section{Value (in)definiteness in EPR type configurations with quantum shares}

Quantum mechanics postulates that a pure state, such as, for example, the entangled singlet Bell state
$\vert \Psi \rangle = \frac{1}{2}\left( \vert +- \rangle - \vert -+ \rangle \right)$,
is the most complete representation of such a quantized system.
Any entangled state such as $\vert \Psi \rangle$ encodes (some degree of) value indefiniteness of
its individual constituents~\cite[\S~10]{schrodinger-gwsidqm2,trimmer,wheeler-Zurek:83}
because of (one-to-one) unitarity,
shifting or rescrambling a state specification to relational properties results in a sort of tradeoff or a zero-sum game of information:
Any (increase of) relational information encoded into a quantum system has to be
compensated by a (decrease of) individual value definiteness of the components of a composite system involved~\cite{zeil-99}.
With $\vert \Psi \rangle$, the value indefiniteness of each one component is maximal,
as there is a 50:50 chance of finding the respective particles in the single-particle states
$\vert - \rangle$ and $\vert + \rangle$ individually, respectively.

We may speculate that $\vert \Psi \rangle$ might be only a macro-state such as in the classical examples mentioned earlier,
and that there are microstates analogous to  $\mathbf{J}$ or  $\mathbf{H}$.
This can be excluded either statistically~\cite{peres222}, or by proof by contradiction~\cite{specker-60,kochen2,cabello-96,pitowsky:218,2015-AnalyticKS}.
These theorems hold relative to the assumptions made; in particular, the simultaneous existence of definite values
reflecting potential (yet counterfactual) experimental outcomes, as well as their independence of the measurement context (or type).

However, if we follow Einstein's original motivation for the EPR paper as explained to Schr\"odinger~\cite{einstei-letter-to-schr,Howard1985171,Meyenn-2011}
this ontologic irreducible randomness (or contextual behavior) on both ends, together with the perfect singlet correlation,
appears mindboggling if not contradictory, to say the least:
because how can the second constituent of the pair ``know'' what outcome the first constituent (possibly together with its local measurement context and the complete
disposition of the measurement instrument) produced?
In particular, if these events or outcomes and their entire local measurement contexts are spatially separated under strict Einstein locality conditions~\cite{wjswz-98}
this question poses a challenge that has no immediate resolution.
(In such a case, the temporal succession of the events or outcomes depends on the reference frame and is, therefore, a matter of conventions
determined by observational means: both could occur simultaneously, or one after another.)
In Einstein's terms~\cite{Howard1985171}, {\em ``The real state of B thus cannot depend upon the kind of measurement I carry out on A.''}
(German underlined original: {``Der wirkliche Zustand von B kann nun nicht davon abbh\"angen, was f\"ur eine Messung ich an A vornehme.''})

Thereby, it is not so much the correlation function which in the quantum case is $\cos \theta$
(as compared to the linear classical correlation $2 \theta / \pi - 1$ already mentioned)
but the specific value indefiniteness of the single seemingly ``isolated'' constituents of the Bell singlet state $\vert \Psi \rangle$
in all spatial directions: if we were to measure, say, a pair of two such (space-like) separated particles ``far (eg, lightyears) away'' from each other,
then regardless of the measurement direction chosen (as long as the directions at both ends are identical), we obtain that
\begin{itemize}

\item[VI] any individual outcome on each of the two ends occurs randomly, reflecting value indefiniteness of individual observables; and yet,

\item[RE] those outcomes are the exact opposite of each other, reflecting the relational encoding of the particle pair.

\end{itemize}

It needs to be stressed that, on each side, there cannot be local contextuality in the sense of Bohr quoted earlier at work---such
at the particular measurement outcome is a reflection of the environment and of the state of the pair.
Indeed, in this singlet state case,
as the individual state of the constituent is value indefinite, only the environment would contribute to the outcome,
and one could expect contextuality to be ``maximal'' or total: because the outcome tells nothing about the state, and is only about the measurement environment.
But how can two supposedly uncorrelated (nonrelational encoded) environments on these two separated ends ``far (say, light years) away'' from each other,
produce such perfect (anti-)correlations on individual, one-by-one pairs of joint outcomes?

Einstein's response was to abandon the completeness of the wave function and go for hidden (parameter) shares.
Another obvious reaction would be to deny spatio-temporal separateness among constituents of an entangled state.

One may, for instance, conjecture, that a wave function like $\vert \Psi \rangle = \frac{1}{2}\left(\vert +- \rangle - \vert -+ \rangle \right)$
corresponds to a macrostate grouping or bundling together $\vert +- \rangle$ and $\vert -+ \rangle $,
and that the microstate, for a particular measurement direction $\boldsymbol{\alpha}$, is either one of the two microstates
$\vert +- \rangle$ or $\vert -+ \rangle $, respectively.
An immediate objection to this would be that this is an ad hoc assumption, and that would require a choice of $\vert +- \rangle$ or $\vert -+ \rangle $
for every direction $\boldsymbol{\alpha}$, resulting in an infinite amount of information in that microstate.
A response to this would be that also in the classical case a precise choice of  $\mathbf{J}$ requires an infinite amount of information.
That is even true for the specification of any quantum observable, in accordance with (but not explicitly mentioned) the
alleged finiteness of the information of a quantum state {\em relative to} a direction~\cite{zeil-99}.


\section{Elastic band toy model for classical local  simulation VI and RE}

\begin{figure}[htb!]
\begin{center}
\begin{tabular}{ccc}
\begin{tikzpicture}  [scale=0.4]

\tikzstyle{every path}=[line width=1pt]


\coordinate (zero) at (0,0);
\coordinate [label={[label distance=2mm]90:$\mathbf{J}_-=-1$}] (sup) at (0,5);
\coordinate [label={[label distance=2mm]270:$\mathbf{J}_+=+1$}]  (sdown) at (0,-5);
\coordinate [label={[label distance=2mm]45:$\mathbf{A}$}]  (a) at (4,3);
\coordinate [label={[label distance=2mm]180:$\mathbf{A}_\mathbf{J}$}]  (a0) at (0,3);
\coordinate [label={[label distance=2mm]0:$x$}]  (x) at (0,-3);

\pic[draw, angle radius=8mm, angle eccentricity=0.6, "$\boldsymbol{\alpha}$"] {angle = a--zero--sup};

\draw[decorate,decoration={coil,segment length=3pt},green] (sdown)--(a0)  node[label={[label distance=0mm]180:{}},pos=0.4, left, black] {$\mathbf{J}\;$};
\draw[decorate,decoration={coil,segment length=3pt},green!100] (a0)--(sup);

\draw[line width=1pt,green,dotted] (a)--(a0);

\draw[line width=2pt,red] (zero)--(a);

\draw[red] (zero) circle(5);

\draw[black,fill=black] (zero) circle(0.15);

\draw[red,fill=white] (a) circle(0.25);

\draw[red,fill=green] (a0) circle(0.25);

\draw[green,fill=green] (sup) circle(0.25);
\draw[green,fill=green] (sdown) circle(0.25);

\draw[black] (x) node {$\bowtie$};

\end{tikzpicture}
\\
(a)
\\
\begin{tikzpicture}  [scale=0.4]

\tikzstyle{every path}=[line width=1pt]


\coordinate (zero) at (0,0);
\coordinate [label={[label distance=2mm]90:$\mathbf{J}_-=-1$}] (sup) at (0,5);
\coordinate [label={[label distance=2mm]270:$\mathbf{J}_+=+1$}]  (sdown) at (0,-5);
\coordinate [label={[label distance=2mm]45:$\mathbf{A}$}]  (a) at (4,3);
\coordinate [label={[label distance=2mm]45:$\mathbf{B}$}]  (b) at (3,4);
\coordinate [label={[label distance=2mm]180:$\mathbf{A}_\mathbf{J}$}]  (a0) at (0,3);
\coordinate [label={[label distance=2mm]180:$\mathbf{B}_\mathbf{J}$}]  (b0) at (0,4);
\coordinate [label={[label distance=2mm]0:$x$}]  (x) at (0,-3);

\pic[draw, angle radius=10mm, angle eccentricity=0.7, "$\boldsymbol{\beta}$"] {angle = b--zero--sup};
\pic[draw, angle radius=15mm, angle eccentricity=0.8, "$\boldsymbol{\theta}$"] {angle = a--zero--b};

\draw[decorate,decoration={coil,segment length=3pt},green] (sdown)--(a0)  node[label={[label distance=0mm]180:{}},pos=0.4, left, black] {$\mathbf{J}\;$};
\draw[decorate,decoration={coil,segment length=3pt},green!100] (a0)--(sup);

\draw[line width=1pt,green,dotted] (a)--(a0);
\draw[line width=1pt,green,dotted] (b)--(b0);

\draw[line width=2pt,red] (zero)--(a);
\draw[line width=2pt,red] (zero)--(b);

\draw[red] (zero) circle(5);

\draw[black,fill=black] (zero) circle(0.15);

\draw[red,fill=white] (a) circle(0.25);
\draw[red,fill=white] (b) circle(0.25);

\draw[red,fill=green] (a0) circle(0.25);
\draw[red,fill=green] (b0) circle(0.25);

\draw[green,fill=green] (sup) circle(0.25);
\draw[green,fill=green] (sdown) circle(0.25);

\draw[black] (x) node {$\bowtie$};

\end{tikzpicture}
\\
(b)
\end{tabular}
\end{center}
\caption{\label{2022-epr-figure1}
(a)
``Inverted'' elastic string model of  Aerts and de Bianchi~\cite{Aerts_2016,Aerts_2014,aerts-69}:
$\mathbf{A}$ stands for the observable located on the unit circle. The state $\mathbf{J}$ is characterized by its angle
(aka position of the sphere, in this drawing it is at angle zero), as well as its
single, unique breaking point $x$.
$\boldsymbol{\alpha}$ is the angle between $\mathbf{J}$ and $\mathbf{A}$.
The ``quantum-type'' cosine law results from the orthogonal projection of $\mathbf{A}$
onto $\mathbf{J}$ at point $\mathbf{A}_\mathbf{J}$, as well as from the assumption that the breaking point $x$ is equidistributed along the line segment
$\overline{\mathbf{J}_+  \mathbf{J}_-}$. Whenever $x$ lies within $\overline{\mathbf{J}_+  \mathbf{A}_\mathbf{J}}$ the observable
$\mathbf{r}_{\boldsymbol{\alpha}}$ is associated with $+1$; otherwise it is $-1$.
(b)
The same elastic string model with two observables $\mathbf{A}$ and $\mathbf{B}$.
}
\end{figure}
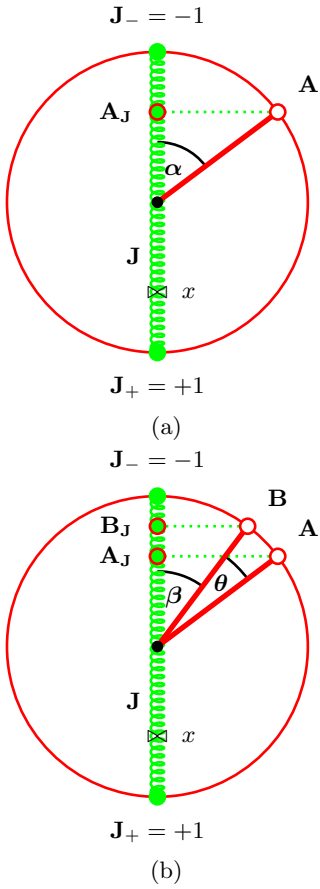

The following should not be understood as a claim about how ``things are'' but rather how one can come up with some classical local construction comforting our
observations on quantized systems.
Let us, for the sake of demonstration, realize a classical toy model that may, under certain circumstances, be considered ``local'',
and realizes both criteria VI and RE mentioned earlier by ``local'' shares.
(The admittedly vague term ``local'' and its circumstances will be specified later.)
It is a modified and ``inverted''---here we exchange the state with the observable, and vice versa;
we also consider the probability of the breaking point instead of the probability of the object moving up or down---elastic sphere model
of Aerts and de Bianchi~\cite{aerts-69,Aerts-91,Aerts_2014,Aerts_2016,aerts-22}.
As depicted in Figure~\ref{2022-epr-figure1}(a) the dichotomic observable $\mathbf{A}$ is characterized by the angle $\boldsymbol{\alpha}$
relative to the angle of the state $\mathbf{J}$.
The state is thought of as an elastic string $\mathbf{J}$, and further characterized by a single, unique breaking point $x$.
For every experiment this breaking point is pre-determined. For multiple experiments, the respective breaking points are evenly distributed over the entire elastic string.
Thus, effectively,
\begin{equation}
A  = \mathrm{sgn} \big( \mathbf{J} \cdot  \mathbf{A} - x \big)
.
\end{equation}

\subsection{Single observable probabilities and expectations}

Since the angle between the observable $\mathbf{A}$ and the elastic string $\mathbf{J}$ is $\boldsymbol{\alpha}$,
and since the breaking point $x$ of the elastic string is supposed to be equidistributed along the line segment $\overline{\mathbf{J}_+  \mathbf{J}_-}$,
and the radius of the unit circle is $1$,
the probability that the breaking point will be observed as lying in-between $\mathbf{J}_+$ and $\mathbf{A}_\mathbf{J}$  is just the length
$\| \overline{\mathbf{J}_+  \mathbf{A}_\mathbf{J}}\| = 1+\cos \boldsymbol{\alpha}$ of the line segment $\overline{\mathbf{J}_+  \mathbf{A}_\mathbf{J}}$,
divided by the length of the diameter 2;
that is,
$P_+(\boldsymbol{\alpha})= \frac{1}{2}\left( 1 + \cos \boldsymbol{\alpha} \right)=\cos^2\frac{\boldsymbol{\alpha}}{2}$.
Likewise, $P_-(\boldsymbol{\alpha})=1-P_+(\boldsymbol{\alpha})=   \frac{1}{2}\left( 1 - \cos \boldsymbol{\alpha} \right)=\sin^2\frac{\boldsymbol{\alpha}}{2}$.
The expectation is an affine transformation of the probabilities; that is,
$E(\boldsymbol{\alpha})=P_+(\boldsymbol{\alpha})-P_-(\boldsymbol{\alpha})= 1 - 2 P_-(\boldsymbol{\alpha})= -1 + 2 P_+(\boldsymbol{\alpha})= \cos \boldsymbol{\alpha}$.

\subsection{Two-observable probabilities and correlations}

EPR-type configurations with elastic band models can be conceptualized by imagining pairs of identical elastic bands that have identical initial states;
that is, both bands would need to be aligned on the same ray, and also their breaking points need to be (inversely) identical.
For a ``singlet'' the orientation of the elastic bands needs to be a relative inverse; that is, imagine two initially identical bands,
with one band rotated $180^\circ$ around its mid-point, so that the directions are essentially inverse.

For two observables $\mathbf{A}$ and $\mathbf{B}$ in a configuration depicted in Figure~\ref{2022-epr-figure1}(b)
with the relative angle $\boldsymbol{\theta} =   \boldsymbol{\beta} -  \boldsymbol{\alpha}$ between the measurement directions
$0\le \boldsymbol{\alpha} \le \pi$ and
$0\le \boldsymbol{\beta} \le   \boldsymbol{\alpha}$
associated with $\mathbf{A}$ and $\mathbf{B}$, respectively,
an analog argument counting the length of the respective line segments on $\mathbf{J}$ yields
\begin{equation}
\begin{split}
P_+( \boldsymbol{\alpha}, \boldsymbol{\beta} ) =  \frac{1}{2}\big[ \left(1 - \cos \boldsymbol{\beta} \right) + \left(1 + \cos \boldsymbol{\alpha} \right) \big]
\\
=
1+  \frac{1}{2}  \left( \cos \boldsymbol{\alpha} - \cos \boldsymbol{\beta} \right)
,\\
P_-( \boldsymbol{\alpha}, \boldsymbol{\beta} ) = \frac{1}{2}  \left( \cos \boldsymbol{\beta} - \cos \boldsymbol{\alpha} \right)
,\\
E( \boldsymbol{\alpha}, \boldsymbol{\beta} ) =  P_+( \boldsymbol{\alpha}, \boldsymbol{\beta} ) - P_-( \boldsymbol{\alpha}, \boldsymbol{\beta} )
\\
 =
1+   \cos \boldsymbol{\alpha} - \cos \boldsymbol{\beta}
\\
=
1+   \cos \boldsymbol{\alpha} - \cos \left( \boldsymbol{\alpha} + \boldsymbol{\theta} \right)
.
\end{split}
\label{2022-epr-efebm3}
\end{equation}
A plausibility check indicates that this correlation function lies in-between $-1$ and $+1$:
for
$\boldsymbol{\alpha} = - \boldsymbol{\theta} =\pi$
and
$\boldsymbol{\beta} =  0$,
$E( \pi , 0 ) 
=  -1$;
likewise,
for
$\boldsymbol{\alpha} = \boldsymbol{\beta}$ and thus $\boldsymbol{\theta} = 0$,
$E( \pi , 0 ) 
= +1$.

\begin{figure}
\begin{center}
\includegraphics[width=0.5\textwidth]{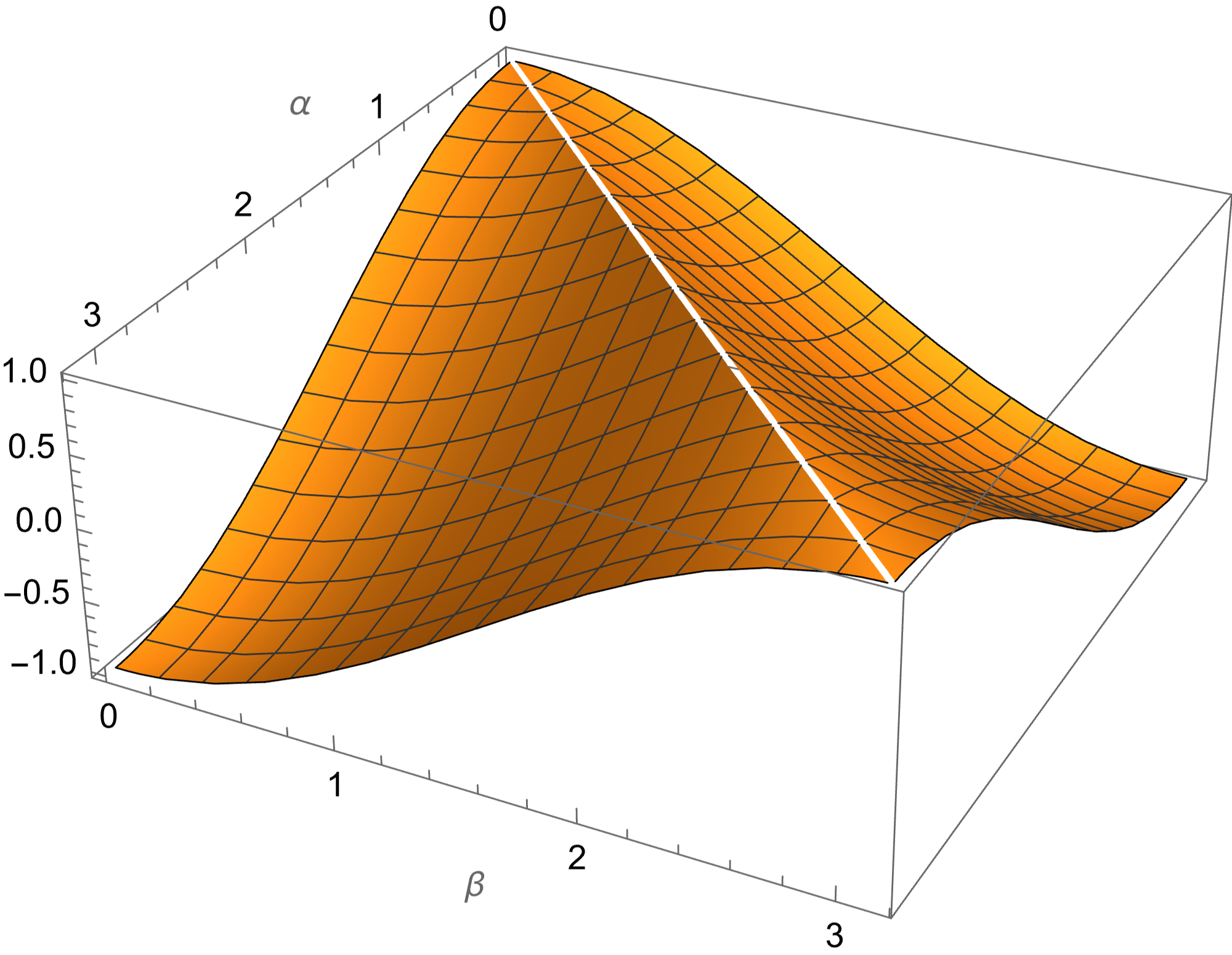}
\end{center}
\caption{Correlation function of the elastic band model.}
\label{2022-epr-Fig1}
\end{figure}

A similar calculation for $\boldsymbol{\beta} \ge   \boldsymbol{\alpha}$ yields the general form for $0 \le  \boldsymbol{\alpha}, \boldsymbol{\beta}   \le \pi$:
\begin{equation}
\begin{split}
E( \boldsymbol{\alpha}, \boldsymbol{\beta})
=
1+   \big( \cos \boldsymbol{\alpha} - \cos  \boldsymbol{\beta}  \big) \mathrm{sgn} \big(\boldsymbol{\alpha} - \boldsymbol{\beta} \big)
\\
=  E( \boldsymbol{\alpha}, \boldsymbol{\alpha} + \boldsymbol{\theta})
=
1-   \big[ \cos \boldsymbol{\alpha} - \cos \left( \boldsymbol{\alpha} + \boldsymbol{\theta} \right) \big]  \mathrm{sgn} \boldsymbol{\theta}
,
\end{split}
\label{2022-epr-efebm5}
\end{equation}
as depicted in Figure~\ref{2022-epr-Fig1}.


Suppose now a protocol in which $\boldsymbol{\alpha}$ is always aligned
along with the share $\mathbf{J}$, such that $\boldsymbol{\alpha}=0$.
Then the correlation in Equation~(\ref{2022-epr-efebm5}) reduces to
\begin{equation}
E( \boldsymbol{\alpha}=0, \boldsymbol{\beta} =   \boldsymbol{\theta} )  \equiv E( \boldsymbol{\theta} )
=
\cos  \boldsymbol{\theta}.
\label{2022-epr-efebm6}
\end{equation}

Another option would be to assume that the direction of $\mathbf{J}$ is uniformly distributed in the interval $[0, \pi ]$, resulting in
\begin{equation}
\begin{split}
E( \boldsymbol{\theta} )
=
\frac{1}{\pi} \int_0^\pi
\big[
1+   \cos \boldsymbol{\alpha} - \cos \left( \boldsymbol{\alpha} - \boldsymbol{\theta} \right)
\big]
d \boldsymbol{\alpha}
\\
= 1  - \frac{2}{\pi} \sin  \boldsymbol{\theta}
.
\end{split}
\label{2022-epr-efebm}
\end{equation}

\subsection{Locality and contextuality}

The quantum-type cosine form of the two-particle expectation function
$E(\mathbf{A},\mathbf{B})$
should give rise to violations of classical Boolean ``conditions of possible experience''~\cite{froissart-81,Pit-94},
in particular, the Clauser-Horn-Shimony-Holt (CHSH) inequality
$-2 \le
E(\boldsymbol{\alpha},\boldsymbol{\beta}) +
E(\boldsymbol{\alpha},\boldsymbol{\beta}') +
E(\boldsymbol{\alpha}',\boldsymbol{\beta}) -
E(\boldsymbol{\alpha}',\boldsymbol{\beta}')
\le 2$.
It is maximally violated~\cite{cirelson:80,filipp-svo-04-qpoly-prl} by quantum mechanics at, for instance,
$\boldsymbol{\alpha}= 0$,
$\boldsymbol{\alpha}'= \pi / 2$,
$\boldsymbol{\beta} = \pi / 4$,
$\boldsymbol{\beta}' = -\pi/4$.
This can be  readily verified by inserting into the quantum expectation functions
$E(\boldsymbol{\alpha},\boldsymbol{\beta}) = \cos ( \boldsymbol{\beta} - \boldsymbol{\alpha} )$,
cosine of the (relative) angles, so that
$
E(0,\pi / 4) +
E(0,-\pi/4) +
E(\pi / 2,\pi / 4) -
E(\pi / 2,-\pi/4)  =
\cos (-\pi/4)+
\cos (\pi/4)+
\cos (\pi/4)-
\cos (3\pi/4)
= 2\sqrt{2}$.

A particular protocol involving the elastic band model shares with quantum mechanics the same type of correlation aka expectation function~(\ref{2022-epr-efebm6})
and should thus also allow violations of the CHSH inequality.
But we just recalled Peres' argument~\cite{peres222} who explicitly demonstrated this by tabulating all classically conceivable configurations; the convex sum of
its entries never violate the CHSH inequality. So how could this happen in the elastic band model nevertheless?

In what follows we shall argue that only an adaptive protocol---adjusting $\boldsymbol{\alpha}$ and $\boldsymbol{\alpha}'$
to coincide with the direction of the share $\mathbf{J}$---can violate the CHSH inequality.
Delayed choice protocols prohibit adaption.
Therefore, if delayed choice under strict Einstein locality conditions is imposed~\cite{wjswz-98,VincentJacques02162007},
we can only employ non-adaptive protocols but not adaptive ones.

First, let us observe that a non-adaptive protocol allowing delayed choice under strict Einstein locality conditions
yielding a correlation function of the type of Equation~(\ref{2022-epr-efebm5}) does not violate the CHSH inequality,
say, for  $\boldsymbol{\alpha}= 0$,
$\boldsymbol{\alpha}= \pi / 2$,
$\boldsymbol{\beta} = \pi / 4$,
$\boldsymbol{\beta}' = -\pi/4$; indeed it renders the value $\sqrt{2}$ for the CHSH sum
$E(\boldsymbol{\alpha},\boldsymbol{\beta}) +
E(\boldsymbol{\alpha},\boldsymbol{\beta}') +
E(\boldsymbol{\alpha}',\boldsymbol{\beta}) -
E(\boldsymbol{\alpha}',\boldsymbol{\beta}')$.
Indeed, for $\boldsymbol{\beta},\boldsymbol{\beta}' \le \boldsymbol{\alpha},\boldsymbol{\alpha}'$ the CHSH sum reduces to
$2 \big( 1 + \cos \boldsymbol{\alpha} - \cos \boldsymbol{\beta} \big) \le 2$ which, as per assumption, $\boldsymbol{\alpha} \ge  \boldsymbol{\beta}$,
is bounded from above and below by the classical bounds.
The second to fifth colums of Table~\ref{2022-epr-T1} enumerate a simulation (similar to Peres' Table~1\cite{peres222})
of all four terms of the CHSH inequality in a non-adaptive setting.

However, if an adaptive protocol like the one yielding~(\ref{2022-epr-efebm6}) is employed,
then the absolute value of the CHSH sum may exceed the maximal classical value 2.
Because in this case the full context---in particular, the necessity to know the choice between either $\mathbf{A}$ or $\mathbf{A}'$
for an unambiguous definition of $\mathbf{B}$ and $\mathbf{B}'$---matters.
The seventh to tenth colums of Table~\ref{2022-epr-T1} enumerate a simulation
of all four terms of the CHSH inequality in an adaptive setting,
based on the same configurations as the simulation of the aforementioned non-adaptive protocol.

\begin{table*}
\caption{\label{2022-epr-T1}
Peres-type valuation table for 20 runs of the elastic string model: the first number indicates the position of the breaking point $x$,
the following group of four numbers enumerates instances of single outcomes
of the observables $\mathbf{A}$, $\mathbf{A}'$, $\mathbf{B}$, and $\mathbf{B}'$,
the second group of five numbers
enumerates the respective expectations
$\mathbf{A}\mathbf{B}$, $\mathbf{A}\mathbf{B}'$, $\mathbf{A}'\mathbf{B}$, and $\mathbf{A}'\mathbf{B}'$
and the resulting CHSH sum
for the non-adaptive protocol,
and the second group of five numbers enumerates these expectations and the resulting CHSH sum for the adaptive protocol.}
\begin{center}
\begin{ruledtabular}
\begin{tabular}{ccccccccccccccc}
&&&&&\multicolumn{5}{l}{non-adaptive, delayed choice}&\multicolumn{5}{l}{adaptive}  \\
$x$&$\mathbf{A}$&$\mathbf{A}'$&$\mathbf{B}$&$\mathbf{B}'$&$\mathbf{A}\mathbf{B}$&$\mathbf{A}\mathbf{B}'$&$\mathbf{A}'\mathbf{B}$&$\mathbf{A}'\mathbf{B}'$&CHSH sum&$\mathbf{A}\mathbf{B}$&$\mathbf{A}\mathbf{B}'$&$\mathbf{A}'\mathbf{B}$&$\mathbf{A}'\mathbf{B}'$&CHSH sum\\
-0.514823 & $+$ & $+$ & $+$ & $+$ & $+$ & $+$ & $+$ & $+$ & 2 & $+$ & $+$ & $+$ & $-$ & 4 \\
-0.832267 & $+$ & $+$ & $+$ & $+$ & $+$ & $+$ & $+$ & $+$ & 2 & $+$ & $+$ & $+$ & $+$ & 2 \\
 0.920526 & $+$ & $-$ & $-$ & $-$ & $-$ & $-$ & $+$ & $+$ & -2 & $-$ & $-$ & $-$ & $-$ & -2 \\
 0.013375 & $+$ & $-$ & $+$ & $+$ & $+$ & $+$ & $-$ & $-$ & 2 & $+$ & $+$ & $+$ & $-$ & 4 \\
 0.444354 & $+$ & $-$ & $+$ & $+$ & $+$ & $+$ & $-$ & $-$ & 2 & $+$ & $+$ & $+$ & $-$ & 4 \\
 0.486249 & $+$ & $-$ & $+$ & $+$ & $+$ & $+$ & $-$ & $-$ & 2 & $+$ & $+$ & $+$ & $-$ & 4 \\
-0.760656 & $+$ & $+$ & $+$ & $+$ & $+$ & $+$ & $+$ & $+$ & 2 & $+$ & $+$ & $+$ & $+$ & 2 \\
 0.425472 & $+$ & $-$ & $+$ & $+$ & $+$ & $+$ & $-$ & $-$ & 2 & $+$ & $+$ & $+$ & $-$ & 4 \\
 0.973582 & $+$ & $-$ & $-$ & $-$ & $-$ & $-$ & $+$ & $+$ & -2 & $-$ & $-$ & $-$ & $-$ & -2 \\
 0.626781 & $+$ & $-$ & $+$ & $+$ & $+$ & $+$ & $-$ & $-$ & 2 & $+$ & $+$ & $+$ & $-$ & 4 \\
-0.35275  & $+$ & $+$ & $+$ & $+$ & $+$ & $+$ & $+$ & $+$ & 2 & $+$ & $+$ & $+$ & $-$ & 4 \\
 0.988427 & $+$ & $-$ & $-$ & $-$ & $-$ & $-$ & $+$ & $+$ & -2 & $-$ & $-$ & $-$ & $-$ & -2 \\
-0.762208 & $+$ & $+$ & $+$ & $+$ & $+$ & $+$ & $+$ & $+$ & 2 & $+$ & $+$ & $+$ & $+$ & 2 \\
 0.735898 & $+$ & $-$ & $-$ & $-$ & $-$ & $-$ & $+$ & $+$ & -2 & $-$ & $-$ & $-$ & $-$ & -2 \\
 0.0588852 & $+$ & $-$ & $+$ & $+$ & $+$ & $+$ & $-$ & $-$ & 2 & $+$ & $+$ & $+$ & $-$ & 4 \\
 -0.498925 & $+$ & $+$ & $+$ & $+$ & $+$ & $+$ & $+$ & $+$ & 2 & $+$ & $+$ & $+$ & $-$ & 4 \\
 -0.53331 & $+$ & $+$ & $+$ & $+$ & $+$ & $+$ & $+$ & $+$ & 2 & $+$ & $+$ & $+$ & $-$ & 4 \\
 -0.822113 & $+$ & $+$ & $+$ & $+$ & $+$ & $+$ & $+$ & $+$ & 2 & $+$ & $+$ & $+$ & $+$ & 2 \\
 0.0398871 & $+$ & $-$ & $+$ & $+$ & $+$ & $+$ & $-$ & $-$ & 2 & $+$ & $+$ & $+$ & $-$ & 4 \\
 -0.226003 & $+$ & $+$ & $+$ & $+$ & $+$ & $+$ & $+$ & $+$ & 2 & $+$ & $+$ & $+$ & $-$ & 4 \\
$\vdots$&$\vdots$&$\vdots$&$\vdots$&$\vdots$& $\vdots$ & $\vdots$ & $\vdots$ & $\vdots$ & $\vdots$ & $\vdots$ & $\vdots$ & $\vdots$ & $\vdots$ & $\vdots$ \\
$\langle x \rangle=0$&&&&&&&&&$\langle \mathrm{CHSH} \rangle=\sqrt{2}$&&&&&$\langle \mathrm{CHSH} \rangle= 2 \sqrt{2}$\\
\end{tabular}
\end{ruledtabular}
\end{center}
\end{table*}

That is, for any evaluation of the correlation functions in the CHSH sum it is not only necessary to know the share
{\it but also to know the particular choice of the context}; that is, in this case,
the choice between  $\mathbf{A}$ or $\mathbf{A}'$, as well as between $\mathbf{B}$ or $\mathbf{B}'$.
Violations of the CHSH inequality require uniformity: the form invariance over all variations states of the (classical) share.
It is insufficient to render, say, the classical cosine form of the quantum correlation function for a particular configuration,
such as setting
$\boldsymbol{\alpha} = 0$
in the general correlation function
$
1+   \cos \boldsymbol{\alpha} - \cos \left( \boldsymbol{\alpha} + \boldsymbol{\theta} \right)
$
of Equation~(\ref{2022-epr-efebm3}), thereby obtaining the quantum $\cos \boldsymbol{\theta}$ form.
Because any other configuration $\boldsymbol{\alpha} \neq 0$ involving shares $\mathbf{J}$ not aligned with $\boldsymbol{\alpha}0$
yields correlations that may ``compensate'' and ``regularize'' the CHSH form to its classical bounds.

\section{Plasticity of the elastic band model}

\begin{figure}[htb!]
\begin{center}
\begin{tabular}{ccc}
\begin{tikzpicture}  [xscale=0.25,yscale=0.4]

\tikzstyle{every path}=[line width=1pt]


\coordinate (zero) at (0,0);
\coordinate [label={[label distance=2mm]90:$\mathbf{J}_-=-1$}] (sup) at (0,5);
\coordinate [label={[label distance=2mm]270:$\mathbf{J}_+=+1$}]  (sdown) at (0,-5);
\coordinate [label={[label distance=2mm]45:$\mathbf{A}$}]  (a) at (4,3);
\coordinate [label={[label distance=2mm]45:$\mathbf{B}$}]  (b) at (3,4);
\coordinate [label={[label distance=1mm]180:$\mathbf{A}_\mathbf{J}$}]  (a0) at (0,3);
\coordinate [label={[label distance=0.5mm]180:$\mathbf{B}_\mathbf{J}$}]  (b0) at (0,4);
\coordinate [label={[label distance=2mm]0:$x$}]  (x) at (0,-3);

\pic[draw, angle radius=10mm, angle eccentricity=0.7, "$\boldsymbol{\beta}$"] {angle = b--zero--sup};
\pic[draw, angle radius=15mm, angle eccentricity=0.8, "$\boldsymbol{\theta}$"] {angle = a--zero--b};

\draw[decorate,decoration={coil,segment length=3pt},green,xscale={0.4/0.25}] (sdown)--(a0)  node[label={[label distance=0mm]180:{}},pos=0.4, left, black] {$\mathbf{J}\;$};
\draw[decorate,decoration={coil,segment length=3pt},green!100,xscale={0.4/0.25}] (a0)--(sup);

\draw[line width=1pt,green,dotted] (a)--(a0);
\draw[line width=1pt,green,dotted] (b)--(b0);

\draw[line width=2pt,red] (zero)--(a);
\draw[line width=2pt,red] (zero)--(b);

\draw[red] (zero) circle(5);

\draw[black,fill=black,xscale={0.4/0.25}] (zero) circle(0.15);

\draw[red,fill=white,xscale={0.4/0.25}] (a) circle(0.25);
\draw[red,fill=white,xscale={0.4/0.25}] (b) circle(0.25);

\draw[red,fill=green,xscale={0.4/0.25}] (a0) circle(0.25);
\draw[red,fill=green,xscale={0.4/0.25}] (b0) circle(0.25);

\draw[green,fill=green,xscale={0.4/0.25}] (sup) circle(0.25);
\draw[green,fill=green,xscale={0.4/0.25}] (sdown) circle(0.25);

\draw[black,xscale={0.4/0.25}] (x) node {$\bowtie$};

\end{tikzpicture}
\\
(a)
\\
\begin{tikzpicture}  [xscale=0.6,yscale=0.4]

\tikzstyle{every path}=[line width=1pt]


\coordinate (zero) at (0,0);
\coordinate [label={[label distance=2mm]90:$\mathbf{J}_-=-1$}] (sup) at (0,5);
\coordinate [label={[label distance=2mm]270:$\mathbf{J}_+=+1$}]  (sdown) at (0,-5);
\coordinate [label={[label distance=2mm]45:$\mathbf{A}$}]  (a) at (4,3);
\coordinate [label={[label distance=2mm]45:$\mathbf{B}$}]  (b) at (3,4);
\coordinate [label={[label distance=2mm]180:$\mathbf{A}_\mathbf{J}$}]  (a0) at (0,3);
\coordinate [label={[label distance=2mm]180:$\mathbf{B}_\mathbf{J}$}]  (b0) at (0,4);
\coordinate [label={[label distance=2mm]0:$x$}]  (x) at (0,-3);

\pic[draw, angle radius=10mm, angle eccentricity=0.7, "$\boldsymbol{\beta}$"] {angle = b--zero--sup};
\pic[draw, angle radius=15mm, angle eccentricity=0.8, "$\boldsymbol{\theta}$"] {angle = a--zero--b};

\draw[decorate,decoration={coil,segment length=3pt},green,xscale={0.4/0.6}] (sdown)--(a0)  node[label={[label distance=0mm]180:{}},pos=0.4, left, black] {$\mathbf{J}\;$};
\draw[decorate,decoration={coil,segment length=3pt},green!100,xscale={0.4/0.6}] (a0)--(sup);

\draw[line width=1pt,green,dotted] (a)--(a0);
\draw[line width=1pt,green,dotted] (b)--(b0);

\draw[line width=2pt,red] (zero)--(a);
\draw[line width=2pt,red] (zero)--(b);

\draw[red] (zero) circle(5);

\draw[black,fill=black,xscale={0.4/0.6}] (zero) circle(0.15);

\draw[red,fill=white,xscale={0.4/0.6}] (a) circle(0.25);
\draw[red,fill=white,xscale={0.4/0.6}] (b) circle(0.25);

\draw[red,fill=green,xscale={0.4/0.6}] (a0) circle(0.25);
\draw[red,fill=green,xscale={0.4/0.6}] (b0) circle(0.25);

\draw[green,fill=green,xscale={0.4/0.6}] (sup) circle(0.25);
\draw[green,fill=green,xscale={0.4/0.6}] (sdown) circle(0.25);

\draw[black,xscale={0.4/0.6}] (x) node {$\bowtie$};

\end{tikzpicture}
\\
(b)
\end{tabular}
\end{center}
\caption{\label{2022-epr-figure2}
``squeezed'' elastic band models, whereby the length of the elastic band is kept constant but the circumference is distorted:
(a) towards weaker-than-quantum correlations;
(b) towards stronger-than-quantum, classical type correlations.
}
\end{figure}
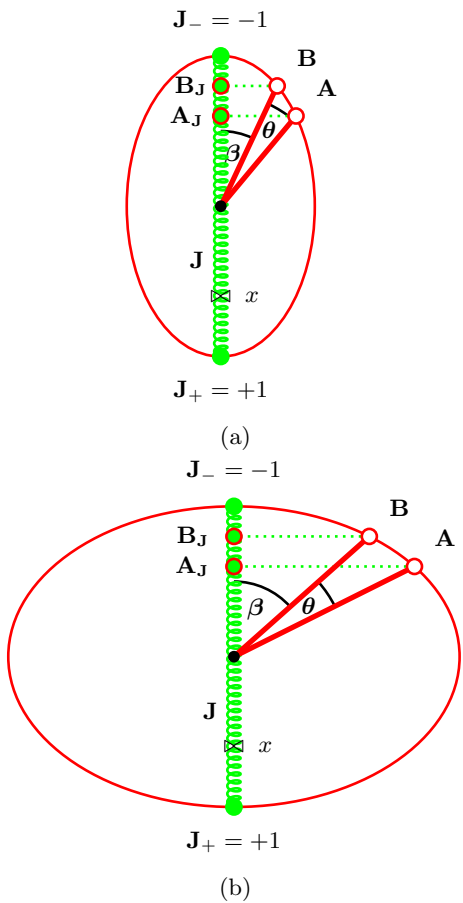

The elastic band model shows some plasticity as we can squeeze or otherwise distort the shape of the circumference
of the circle in Figure~\ref{2022-epr-figure1}.
For the sake of a shape distortion, consider elliptic shapes as drawn in Figure~\ref{2022-epr-figure2} resulting from a
``squeezing'' of the outer circle, whereby the length of the elastic band is kept constant.
The length of the convex shape changes under such squeezing.
Therefore, the parametrization in terms of the length of the outer shape needs to be compensated and renormalized.
The model suggests that there exist two limits: one classical limit obtained by increasing the eccentricity
through decreasing the minor axis in Figure~\ref{2022-epr-figure2}(a),
and one limit yielding a unit step function~\cite{svozil-krenn} centered around the mid-point of the elastic band
by increasing the eccentricity
through increasing the major axis in Figure~\ref{2022-epr-figure2}(b).
Similar distortion transformations have (without a concrete model) been discussed in a previous manuscript~\cite{svozil-2001-cesena}.

\section{Discussion and outlook}

The general goal of this paper was twofold:
first, to point out that it is not sufficient to recover, by classical means, a quantum-type correlation function or some non-linear, trigonometric
(for instance, cosine) form of probability of elementary propositions for a full reconstruction of quantum predictions.
And second, more important, it has been argued that the mindboggling features of quantum mechanics reveal themselves only
through delayed choice measurements under strict Einstein locality conditions.

Because as long as we do not enforce delayed choice under strict Einstein locality conditions
we can have epistemic randomness as well as total relational dependence of a pair of observables
following the quantum predictions.
To construct such a classical model, we modified an elastic band model of Aerts and computed the respective quantum type
probabilities and correlation functions of ``singlet'' states under certain assumptions, in particular (non-)adaptive measurements.

Such a model could perfectly mimic quantum predictions in the usual setup. For instance, if the four terms in the CHSH sum are measured one after the other,
such that the respective contexts are well known to the two observers
by allowing a communication about which context is used; or
alternatively, measuring the terms entering the CHSH sum consecutively, that is, one after the other in a coordinated fashion---for instance,
by performing observations
``$\mathbf{A}\mathbf{B}$'' after breakfast,
$\mathbf{A}\mathbf{B}'$ after lunch,
$\mathbf{A}'\mathbf{B}$ after teatime, and
$\mathbf{A}'\mathbf{B}'$ after supper''---the elastic band model can deliver quantum performance.
Indeed, by deforming the circumference of the elastic band we obtain a wide variety of correlation functions;
and even stronger-than classical correlations such as approximations to the unit step function~\cite{svozil-krenn}.

At first glance, this appears not dissimilar to protocols invoking the transfer of a single bit~\cite{toner-bacon-03,svozil-2004-brainteaser}.
However, there is one decisive difference: whereas the previous protocols quoted require
the direct communication of actual respective measurement outcomes between the parties,
the adaptive protocol introduced here for the elastic band model just requires communication of the context.
In the CHSH simulation case, this is a single bit.
(In that aspect, our protocol is not dissimilar to the Wiesner~\cite{wiesner} and BB84~\cite{benn-84} schemes requiring classically communicating the choice of basis.)
If this bit is denoted by $1 \text{ co-bit}$ then it is ``weaker'' than $1 \text{ bit} (supraluminal communication)$,
that is,  $1 \text{ co-bit}  \prec 1 \text{ bit}$, because it would be possible to simulate a
(Popescu-Rohrlich) non-local machine~\cite{popescu-2014}
because one can simulate the latter without communication.
Moreover, a $1 \text{ co-bit}$ can be used to simulate a (Popescu-Rohrlich) non-local machine and the associated $1 \text{ nl-bit}$
by identifying it with the hidden variable $\lambda$ transferred~\cite{cerf-gisin-massar-pop-04}.

While it has been pointed out that
the direct communication of actual respective measurement outcomes can be replaced by the invocation of a non-local resource~\cite{cerf-gisin-massar-pop-04}
one may still ask whether the only way to realize such a non-local resource is by means of a concealed internal signal
between its ports~\cite{AertsSven}.
The protocol introduced by Cerf, Gisin, Massar, and Popescu~\cite{cerf-gisin-massar-pop-04} invoking a generic (Popescu-Rohrlich) non-local machine,
in combination with the realization of such a (Popescu-Rohrlich type) non-local machine (that uses non-local rubber band shares and pulls and remains causal such that no superluminal
signalling occurs) by Sven Aerts~\cite{AertsSven} is capable of delivering a violation of the CHSH inequality by classical means,
similar to the direct communication of the context discussed earlier.
We conjecture that the protocol introduced by Cerf, Gisin, Massar, and Popescu, augmented by Aerts' non-local resources,
can be generalized to render stronger-than-classical violations of the CHSH inequality.

We also conjecture that, in analogy to the elastic band model discussed,
if the non-local resource is required to be perform uniformly over all variations of legal states of the share,
this goal cannot be achieved by any strictly local means.

One might say that because the information in an entangled (say, a singlet) state
of two constituents appears to be encoded purely by relational sampling~\cite{zeil-99,svozil-2016-sampling},
quantum entanglement cannot provide any means to encode any, possibly hidden, internal share $\mathbf{J}$.
Insofar as the respective outcomes are uncontrollable this is consistent with relativity theory, as no faster-than-light signalling can be achieved.

The analogy between such types of ``non-local'' signalling and the cloning (aka copying) of bits is intriguing:
while this kind of signalling seems to be allowed for one parameter setting only~\cite{AertsSven},
so is the copying of a fixed single bit~\cite[Eq.~(2.12), page~40]{mermin-07}.

This has far-reaching consequences for all forms of ``quantum certifications'' in security applications,
such as the production of random bits certified by value indefiniteness~\cite{svozil-2009-howto,10.1038/nature09008,Abbott:2010uq,Trejo2020Aug},
or EPR-based quantum cryptography~\cite{ekert91}.
The implementations of all such protocols are only ``good'' relative to the means (not) guaranteed; in particular, strict Einstein locality.

Let us once more contemplate the bigger picture and review the strategy pursued here:
One way of ``explanation'' of the singlet-type behavior on both sides of the EPR-type configuration
is to assume that the respective constituents of a pair carry a common share that determines the outcomes.
Those outcomes appear random because the shares are assumed to be sampled randomly.
In such a conception randomness resides not in the measurement that supposedly ``reveals'' aspects or properties of the share,
but in the randomized shares.

One straightforward way of doing this is Peres' bomb model mentioned earlier.
However, this model cannot deliver violations of, say, the CHSH inequality.
In order to explain a violation of Bell-type inequalities---Boole's conditions of possible experience---a price has to be paid.
One token would be one or more bits of communication between the parties at both ends of the EPR configuration:
either by informing the other side of one's outcome(s)~\cite{toner-bacon-03,svozil-2004-brainteaser},
or by informing the other side of (part of) the measurement context---a method proposed here.
Still another possibility would be to supply the parties with a non-local machine~\cite{cerf-gisin-massar-pop-04,AertsSven}.

My present thinking---as tentative and highly speculative and hypothetical as it seems---is that we are living in a vector world,
and those entangled quantum states are not really spatio-temporally separated at all.
A pure state, encoded by a vector, {\it is} the share; and a complete one at that:
this is all the observers have got. There is no ``hidden'' share, such as those discussed earlier, triggering their outcomes.
Both observers access and share one and the same vector, a formalization of a pure entangled state.
As long as peaceful existence ensues this might not necessarily imply a revision of the construction of space-time coordinate frames.
We shall study such a scenario in a second part of this series.

\ifx\revtex\undefined

\funding{This research was funded in whole, or in part, by the Austrian Science Fund (FWF), Project No. I 4579-N. For the purpose of open access, the author has applied a CC BY public copyright licence to any Author Accepted Manuscript version arising from this submission.
}


\conflictsofinterest{

The author declares no conflict of interest.
The funders had no role in the design of the study; in the collection, analyses, or interpretation of data; in the writing of the manuscript, or in the decision to publish the~results.}

\else

\begin{acknowledgments}

I kindly acknowledge discussions with and suggestions by Diederik Aerts and Sven Aerts.

This research was funded in whole, or in part, by the Austrian Science Fund (FWF), Project No. I 4579-N. For the purpose of open access, the author has applied a CC BY public copyright licence to any Author Accepted Manuscript version arising from this submission.

The author declares no conflict of interest.
\end{acknowledgments}

\fi

\ifx\revtex\undefined

\end{paracol}
\reftitle{References}


 \externalbibliography{yes}
 \bibliography{svozil,ufo}

\else


%

\fi
\end{document}